\documentstyle[seceq,epsfig]{ptptex}

\def\be{\begin{equation}}
\def\ee{\end{equation}}
\def\ulap{\underline\Delta}
\def\unab{{\overline\nabla}}

\title{
Darwin-Riemann problems in general relativity
}

\author{
Silvano {\sc Bonazzola},
Eric {\sc Gourgoulhon},
Philippe {\sc Grandcl\'ement} 
and Jean-Alain {\sc Marck}
}

\inst{
D\'epartement d'Astrophysique Relativiste et de Cosmologie, UMR~8629 du
CNRS, Observatoire de Paris, 92195 Meudon Cedex, France
}

\recdate{
}

\abst{
A review is given of recent results about the computation of
irrotational Darwin-Riemann configurations in general relativity. 
Such configurations are expected to represent fairly well the late stages
of inspiralling binary neutron stars. 
}

\begin{document}

\maketitle

\section{Introduction} \label{s:intro}

The classical Darwin ellipsoids \cite{Chand69} are equilibrium figures of
incompressible fluid bodies in a synchronous binary system\footnote{Contrary
to MacLaurin ellipsoids, which are {\em exact} solutions for rotating 
incompressible fluids in Newtonian gravity, 
Darwin ellipsoids are {\em approximate} solutions, 
because of the second order truncation in the expansion of the 
gravitational potential of the companion.}.
{\em Synchronous} means that each body is spinning with respect to some
inertial frame at the same angular velocity as the orbital angular velocity.
In this manner it always presents the same face to its companion. 
The vorticity of the fluid with respect to some inertial frame is then
equal to twice the orbital angular velocity. 
Darwin-Riemann configurations \cite{LaiRS94} are generalizations of Darwin
ellipsoids to arbitrary vorticity (i.e. non-synchronous spins). 
As detailed below, a subset of Darwin-Riemann ellipsoids is of particular
importance for the late stages of inspiralling binary neutron stars, 
which are expected to be one of the strongest sources of gravitational
radiation for the interferometric detectors currently 
under construction (GEO600, LIGO, TAMA and VIRGO). This subset is 
the {\em irrotational} Darwin-Riemann configurations, i.e. configurations
for which the fluid vorticity with respect to some inertial frame 
vanishes identically. The fluid motion with respect to the inertial frame
is then more or less a circular translation, whereas in a frame
which follows the orbital motion (designed hereafter as the
 {\em co-orbiting frame}), it is a counter-rotation. 
For a more extensive description of irrotational Darwin-Riemann configurations, 
we report to Eriguchi's review \cite{Erigu99}.

The present article focuses on
the general relativistic treatment of irrotational binary configurations.
These configurations can be seen as generalizations
of the irrotational Darwin-Riemann ellipsoids in the following directions:
\begin{enumerate}
\item the fluid is no longer assumed to be incompressible;
\item the gravitational potential of the companion is no longer truncated to
      the second order, but totally considered;
\item the Newtonian treatment is replaced by a general relativistic one.
\end{enumerate}
The motivation for such a study is twofold:
\begin{enumerate}
\item {\bf Investigate the stability with respect to gravitational collapse:}
by means of numerical computations, Wilson, Mathews and 
Marronetti\cite{WilsoM95,WilsoMM96} have found that, due to relativistic effects, binary neutron 
stars may individually collapse to black hole prior to merger. 
This rather surprising result is now thought to be due to an error in some
analytical formula implemented in the numerical code \cite{Flana99}. 
Consequently this motivation is now rather weak. 
\item {\bf Provide realistic initial conditions for binary coalescence:}
it has been shown that the gravitational-radiation driven evolution of a neutron
star binary system is
too rapid for the viscous forces to synchronize the spin of each
neutron star with the orbit \cite{Kocha92,BildsC92} as they do for
ordinary stellar binaries.  Rather, the viscosity is negligible and 
the fluid velocity circulation (with respect to some inertial
frame) is conserved in these systems.  Provided that the initial spins
are not in the millisecond regime, this means that close binary
configurations are better approximated by zero vorticity (i.e. {\em
irrotational}) states than by synchronized states. These irrotational
configurations constitute realistic initial conditions for fully 
hydrodynamical computations of the merging phase, as performed
by different groups \cite{Suen99,Oohara99,Shiba99}.
\end{enumerate}

The plan of this article is as follows. Having presented the general
formalism to treat relativistically irrotational binary systems in 
Sect.~\ref{s:formalism}, we specialize it to the case where the 
spatial 3-metric is assumed to be conformally flat in Sect.~\ref{s:conf_flat},
and exhibit the full system of partial differential equations to be
solved. Some analytical solutions are presented in 
Sect.~\ref{s:analytic}. We then discuss numerical techniques to 
solve the problem in Sect.~\ref{s:num}, and present the numerical
results obtained by various groups in Sect.~\ref{s:num_res}. The paper
ends with a discussion about the innermost stable circular orbit in these
numerical solutions (Sect.~\ref{s:ISCO}). 

\section{Formalism for relativistic irrotational binaries}
\label{s:formalism}

Generalizing the Newtonian formulation presented in Ref.~\citen{BonazGHM92}
we have proposed a relativistic formulation for quasiequilibrium
irrotational binaries \cite{BonazGM97b}.  
The method is
based on one aspect of irrotational motion, namely the {\em
counter-rotation} (as measured in the co-orbiting frame) of the fluid
with respect to the orbital motion.
This formulation has been slightly corrected by 
Asada \cite{Asada98} who has shown that in order to lead unambiguously
to a counter-rotating state, the iterative procedure presented in 
Ref.~\citen{BonazGM97b} must be initiated with a vanishing velocity field
with respect to the co-orbiting observer. Asada \cite{Asada98} 
has also shown that the relativistic definition of counter-rotation 
implies that the flow is irrotational, i.e. that the vorticity 2-form
vanishes identically. 

Subsequently, Teukolsky \cite{Teuko98} and Shibata \cite{Shiba98} gave 
independently two
formulations based on the very definition of irrotationality, which implies
that the specific enthalpy times the fluid 4-velocity is the gradient
of some scalar field \cite{LandaL89} ({\em potential flow}).  

The formulations presented by us \cite{BonazGM97b} (as amended by 
Asada \cite{Asada98}), Teukolsky \cite{Teuko98} and Shibata \cite{Shiba98} 
are equivalent; however the one given by Teukolsky and by
Shibata greatly simplifies the problem. Consequently we used it in the
following discussion. 

The general relativistic treatment of irrotational binary systems is based
on two physically well justified assumptions: (i) quasiequilibrium state (which
means a steady state in the co-orbiting frame), and (ii) irrotational flow. 
Let us examine successively these two assumptions and their relativistic
(geometrical) translation. 

\subsection{Quasiequilibrium assumption}

\begin{figure}
\centerline{ \epsfig{figure=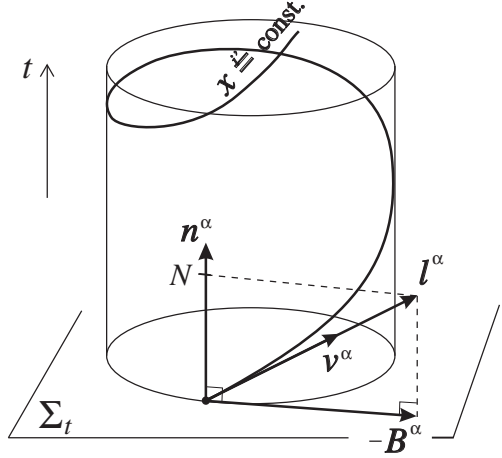,height=6cm} }
\vspace{1cm}
\caption[]{\label{f:heli}   
Spacetime foliation $\Sigma_t$, helicoidal Killing vector
$\mib{l}$ and its trajectories $x^{i'} = {\rm const}$, which are the 
worldlines of the co-orbiting observer (4-velocity: $\mib{v}$).
Also shown are the rotating-coordinate shift vector $\mib{B}$
and the unit future-directed vector $\mib{n}$, normal to the spacelike
hypersurface $\Sigma_t$.} 
\end{figure}

When the separation between the centres of the two neutron stars
is about $50{\rm\, km}$ (in harmonic coordinates)
the time variation of the orbital period, $\dot P_{\rm orb}$, computed at 
the 2nd Post-Newtonian (PN) order by means of the formulas established by 
Blanchet et al. \cite{BlancDIWW95} is about $2\%$.
The evolution at this stage can thus be still considered as a sequence
of equilibrium configurations.
Moreover the orbits are expected to be circular (vanishing eccentricity), 
as a consequence of the gravitational radiation reaction \cite{Peter64}.
In terms of the spacetime geometry, 
we translate these assumptions by
demanding that there exists a Killing vector field $\mib{l}$ which is
expressible as \cite{BonazGM97b}
\be \label{e:helicoidal}
 \mib{l} = \mib{k} + \Omega \, \mib{m} \ ,
\ee
where $\Omega$ is a constant, to be identified with the orbital
angular velocity with respect to a distant inertial observer, and
$\mib{k}$ and $\mib{m}$ are two vector fields with the following 
properties. $\mib{k}$ is timelike at least far from the binary and 
is normalized so that far from the star it coincides
with the 4-velocity of inertial observers. $\mib{m}$ is spacelike,
has closed orbits, is zero on a two dimensional timelike surface, called
the {\em rotation axis}, and is normalized so that 
$\nabla(\mib{m}\cdot\mib{m}) \cdot \nabla(\mib{m}\cdot\mib{m}) / 
(4 \, \mib{m}\cdot\mib{m})$
tends to $1$ on the rotation axis [this latter condition ensures that
the parameter $\varphi$ associated with $\mib{m}$ along its trajectories
by $\mib{m} = \partial/\partial \varphi$ has the standard $2\pi$
periodicity].
Let us call $\mib{l}$ the {\em helicoidal Killing vector}. We assume
that $\mib{l}$ is a symmetry generator not only for the spacetime
metric $\mib{g}$ but also for all the matter fields. In particular,
$\mib{l}$ is tangent to the world tubes representing the surface of 
each star, hence its qualification of {\em helicoidal} 
(cf. Figure~\ref{f:heli}).

The approximation suggested above amounts to neglect outgoing gravitational
radiation. For non-axisymmetric systems --- as binaries are ---
imposing $\mib{l}$ as an exact 
Killing vector leads to a spacetime which is not asymptotically flat
\cite{GibboS83}. Thus, in solving for the gravitational field equations,
a certain approximation has to be devised in order to 
avoid the divergence of some metric coefficients at infinity. For instance
such an approximation could be the Wilson \& Mathews scheme \cite{WilsM89}
that amounts to solve only for the Hamiltonian and momentum constraint 
equations (cf. Sect.~\ref{s:conf_flat}). 
This approximation has been used in all the relativistic studies to date 
and is consistent 
with the existence of the helicoidal Killing vector field (\ref{e:helicoidal}).
Note also that since the gravitational radiation reaction shows up
only at the 2.5-PN order, the helicoidal symmetry is exact up to the
2-PN order.

Following the standard 3+1 formalism, we introduce a spacetime foliation 
by a family of spacelike hypersurfaces $\Sigma_t$ such that at spatial 
infinity, the vector $\mib{k}$ introduced in Eq.~(\ref{e:helicoidal})
is normal to $\Sigma_t$ and the ADM 3-momentum in $\Sigma_t$ vanishes 
(i.e. the time $t$ is the proper time of an asymptotic inertial observer
at rest with respect to the binary system). Asymptotically, 
$\mib{k} = \partial/\partial t$ and $\mib{m} = \partial/\partial\varphi$,
where $\varphi$ is the azimuthal coordinate associated with the above
asymptotic inertial observer, so that Eq.~(\ref{e:helicoidal}) can be
re-written as
\be
 \mib{l} = {\partial\over\partial t} 
		+ \Omega \, {\partial\over\partial \varphi} \ .
\ee

One can then introduce the 
shift vector $\mib{B}$ of co-orbiting coordinates by means of the 
orthogonal decomposition of $\mib{l}$ with respect to the $\Sigma_t$
foliation (cf. Fig.~\ref{f:heli}):
\be
	\mib{l} = N \, \mib{n} - \mib{B} ,
\ee 
$\mib{n}$ being the unit future directed vector normal to $\Sigma_t$ and $N$
the lapse function. 

\subsection{Irrotational flow}

We consider a perfect fluid, which constitutes an excellent approximation 
for neutron star matter. The matter stress energy tensor is then
\be
   \mib{T} = (e+p) \mib{u} \otimes \mib{u} + p \mib{g} \ , 
\ee
$e$ being the fluid proper energy density, $p$ the fluid pressure and
$\mib{u}$ the fluid 4-velocity. 
A zero-temperature equation of state (EOS) is a very good approximation for 
neutron star matter. For such an EOS, the first law of 
thermodynamics gives rise to the following identity (Gibbs-Duhem relation):
\be \label{e:Gibbs-Duhem}
	{\nabla p \over e+p} = {1\over h} \nabla h \ ,
\ee
where $h$ is the fluid specific enthalpy:
\be
	h := {e+p \over m_{\rm B} n} ,
\ee
$n$ being the fluid baryon number density and $m_{\rm B}$ the mean baryon 
mass: $m_{\rm B} = 1.66\times 10^{-27} {\rm\ kg}$. 
Note that for our zero-temperature EOS, $m_{\rm B} \, h$ is equal to the
fluid chemical potential.

By means of the identity (\ref{e:Gibbs-Duhem}), it is straightforward to 
show that the classical momentum-energy conservation equation 
$\nabla\cdot \mib{T} = 0$ is equivalent to the set of two 
equations\cite{Lichn67,Carte79}:
\be \label{e:canon}
	\mib{u} \cdot (\nabla \wedge \mib{w}) = 0 \ ,
\ee
\be \label{e:baryon_conserv}
	\nabla\cdot (n \mib{u}) = 0 \ . 
\ee
In Eq.~(\ref{e:canon}), $\mib{w}$ is the co-momentum 1-form
\be
	\mib{w} := h \, \mib{u}
\ee
and $\nabla \wedge \mib{w}$ denotes the exterior derivative of $\mib{w}$, 
i.e. the vorticity 2-form\cite{Lichn67}. In terms of components, one has
\be
	(\nabla \wedge \mib{w})_{\alpha\beta} = \nabla_\alpha w_\beta
	- \nabla_\beta w_\alpha \ . 
\ee
The advantage of writing the equation of motion in the form (\ref{e:canon})
rather than in the traditional form $\nabla\cdot \mib{T} = 0$ is that
one can see immediately that a flow of the form
\be \label{e:potential_flow}
	\mib{w} = \nabla \Psi \ , 
\ee
where $\Psi$ is some scalar, is a solution of the equation of motion. 
Such a flow is called a {\em potential flow}. Indeed, 
Eq.~(\ref{e:potential_flow}) implies the vanishing of the vorticity 2-form:
\be \label{e:irrot}
	\nabla \wedge \mib{w} = 0 \ ,
\ee
so that the equation of motion (\ref{e:canon}) is trivially satisfied. 
Equation~(\ref{e:irrot}) is the relativistic definition of an irrotational 
flow \cite{Lichn67}.

\subsection{First integral of motion}

The above two assumptions, namely (i) $\mib{l}$ is a symmetry generator and
(ii) the flow is irrotational, yield to the following first integral
\be \label{e:int_prem}
	h \, \mib{l}\cdot \mib{u} = {\rm const}. 
\ee
This was first pointed out by Carter \cite{Carte79}. The demonstration is
straightforward if one applies the classical Cartan's identity to the
1-form $\mib{w}$ and the vector field $\mib{l}$:
\be \label{e:Cartan}
	\pounds_{\mib{l}} \mib{w} = \mib{l}\cdot(\nabla\wedge\mib{w})
	+ \nabla(\mib{l}\cdot\mib{w}) \ , 
\ee
where $\pounds_{\mib{l}}$ denotes the Lie derivative along the vector field
$\mib{l}$. 
Hypothesis (i) implies that $\pounds_{\mib{l}} \mib{w} = 0$ and hypothesis
(ii) that $\nabla \wedge \mib{w} = 0$, so that Eq.~(\ref{e:Cartan}) reduces to
\be
    \nabla( \mib{l} \cdot \mib{w} ) = 0 \ , 
\ee
from which the first integral (\ref{e:int_prem}) follows. 

\subsection{Remark about Bernoulli's theorem}

Let us mention a point which seems to have been missed by various authors:
the existence of the first integral of motion (\ref{e:int_prem}) is not 
merely the relativistic generalization of Bernoulli's theorem. 
This latter follows solely from the existence of the symmetry 
generator $\mib{l}$ and can be established as follows. Inserting
$\pounds_{\mib{l}} \mib{w} = 0$ into Cartan's identity (\ref{e:Cartan}) yields
\be
   \mib{l}\cdot(\nabla\wedge\mib{w})
	+ \nabla(\mib{l}\cdot\mib{w}) = 0 \ . 
\ee
Performing the scalar product by $\mib{u}$ leads to
\be
   \mib{l}\cdot(\nabla\wedge\mib{w})\cdot \mib{u}
	+ \mib{u}\cdot\nabla(\mib{l}\cdot\mib{w}) = 0 \ . 
\ee
The first term in the left-hand side vanishes by virtue of the equation 
of motion (\ref{e:canon}), so that one is left with
\be
     \mib{u}\cdot\nabla(\mib{l}\cdot\mib{w}) = 0 \ , 
\ee
which means that the quantity $\mib{l}\cdot\mib{w} = h \, \mib{l}\cdot \mib{u}$
is constant {\em along each streamline}. This is the Bernoulli theorem. 
The key point is that, in order for the
constant to be uniform over the streamlines (i.e. to be a constant over
spacetime), as in Eq.~(\ref{e:int_prem}), some additional property of the
flow must be required. One well known possibility is {\em rigidity} (i.e.
$\mib{u}$ colinear to $\mib{l}$) \cite{Boyer65}, which would apply 
to synchronized binaries. The alternative property
with which we are concerned here is {\em irrotationality}.

\subsection{Equation for the velocity potential $\Psi$}

Since Eq.~(\ref{e:canon}) is trivially satisfied by the potential flow
(\ref{e:potential_flow}), the only part of the momentum-energy equation 
$\nabla\cdot\mib{T}=0$ which remains to be satisfied is 
Eq.~(\ref{e:baryon_conserv}) (baryon number conservation). 
Inserting Eq.~(\ref{e:potential_flow}) in Eq.~(\ref{e:baryon_conserv})
results in an equation for the
velocity potential $\Psi$:
\be \label{e:eq_Psi_4d}
{n \over h} \nabla \cdot \nabla \Psi 
        + \nabla \Psi \cdot \nabla \left( {n \over h} \right) =0 \ .
\ee

Following the 3+1 formalism, we introduce the 3-metric $\mib{h}$ induced
by $\mib{g}$ into the $\Sigma_t$ hypersurfaces, and denote by $D$ the
associated covariant derivative.  Taking into account the helicoidal 
symmetry, Eq.~(\ref{e:eq_Psi_4d}) becomes\footnote{Latin indices $i,j,\ldots$ 
run from 1 to 3.}
\be \label{e:psicov}
n D_i D^i \Psi + D^i n \,  D_i \Psi 
  = - {h \Gamma_{\rm n} \over N} B^i D_i n
  + n \left(  D^i \Psi \, D_i \ln {h\over N}
     - {B^i \over N} D_i \Gamma_{\rm n} \right)
  + K n h \Gamma_{\rm n}
\ ,
\ee
where K is the
trace of the extrinsic curvature tensor of the $\Sigma_t$ hypersurfaces
and $\Gamma_{\rm n}$ is the Lorentz factor between the fluid observer and
the Eulerian observer (observer whose 4-velocity is the unit normal $\mib{n}$
to $\Sigma_t$): 
\be \label{e:gamma_n}
\Gamma_{\rm n} := - \mib{n}\cdot\mib{u}
	= \left( 1 + {1\over h^2} 
                D_i \Psi \, D^i \Psi \right) ^{1/2} \ . 
\ee

\section{A simplifying assumption: the conformally flat 3-metric}
\label{s:conf_flat}

\subsection{Presentation and justifications} \label{s:justif}

The problem of finding irrotational binary configurations in quasiequilibrium
amounts to solve the elliptic equation (\ref{e:psicov}) for $\Psi$ and 
the Einstein equations for the components of the metric tensor. 
As a first step, a simplifying assumption can be introduced, in
order to reduce the computational task, namely to take the 3-metric 
induced in the hypersurfaces $\Sigma_t$ to be conformally flat:
\be \label{e:conf_plat}
	\mib{h} = A^2 \, \mib{f} \ , 
\ee
where $A$ is some scalar field and $\mib{f}$ is a flat 3-metric. 
Note that this assumption, first introduced by Wilson \& Mathews \cite{WilsM89},
is physically less justified than the assumptions of quasiequilibrium and
irrotational flow discussed above. However, some possible justifications of 
(\ref{e:conf_plat}) are
\begin{enumerate}
\item it is exact for spherically symmetric configurations;
\item it is very accurate for axisymmetric rotating neutron 
stars \cite{CookST96};
\item the 1-PN metric fits it;
\item the 2.5-PN metric deviates from it by only 2 \% 
for two $1.4 \, M_\odot$ 
neutron stars as close as
30~km (in harmonic coordinates) \cite{BonazGM99c}.
\end{enumerate}

\subsection{Partial differential equations for the metric}

Assuming (\ref{e:conf_plat}), the full spacetime metric takes the 
form 
\begin{equation} \label{e:met}
ds^2 = -(N^2 - B_i B^i) dt^2 - 2 B_i dt\, dx^i 
        + A^2 f_{ij} dx^i\, dx^j \ .
\end{equation}
We have thus five metric functions to determine: the lapse $N$, the
conformal factor $A$ and the three components $B^i$ of the shift vector. 
Let us introduce auxiliary metric quantities: the shift vector
of non-rotating coordinates:
\be \label{e:shift_non_rot}
	\mib{N} = \mib{B} + \Omega {\partial\over\partial \varphi} \ , 
\ee
and the two logarithms
\be
	\beta := \ln(AN) , 
\ee
\be
	\nu := \ln N \ . 
\ee
At the Newtonian limit $\beta=0$ and $\nu$ coincides with the Newtonian
gravitational potential. 

In the following, we choose maximal slicing coordinates, for which $K=0$. 

The Killing equation $\nabla_\alpha l_\beta + \nabla_\beta l_\alpha = 0$
give rise to a relation between the $\Sigma_t$ extrinsic curvature tensor
$\mib{K}$ and the shift vector $\mib{N}$:
\be \label{e:kij}
K^{ij} = - {1 \over 2 A^2 N}
          \left\{
                \unab^i N^j + \unab^j N^i
                  - {2 \over 3} f^{ij} \unab_k N^k 
          \right\} \ ,
\ee
where $\unab$ stands for the covariant derivative associated with the flat
3-metric $\mib{f}$. 

The trace of the spatial part of the Einstein equation, combined with the
Hamiltonian constraint equation, result in the following two equations
\be \label{e:beta}
\ulap \beta = 4 \pi A^2 S + {3 \over 4} A^2 K_{ij} K^{ij} 
        - {1 \over 2} \left( \unab_i \nu \unab^i \nu  
        		+ \unab_i \beta \unab^i \beta \right) \ ,
\ee
\be \label{e:nu}
\ulap \nu = 4 \pi A^2 (E + S) + A^2 K_{ij} K^{ij} - \unab_i \nu \unab^i \beta 
   \ ,
\ee
where $\ulap := \unab^i \unab_i$ is the Laplacian operator associated with 
the flat metric $\mib{f}$, $E$ and $S$ are respectively the matter energy 
density and trace of the stress tensor, both as measured 
by the Eulerian observer:
\be
        E := \mib{n}\cdot\mib{T}\cdot\mib{n} = \Gamma_{\rm n}^2 (e+p) - p,
\ee
\be
	S := \mib{h}\cdot\mib{T} = 3p +  (E +p) U_i U^i \ , 
\ee
$U^i$ being the fluid 3-velocity as measured by the Eulerian observer:
\be
	\mib{U} := {1\over \Gamma_{\rm n}} \, \mib{h}\cdot\mib{u} \ . 
\ee
For
the potential flow (\ref{e:potential_flow}), $U^i$ is related to $\Psi$ by
\be
	U^i = {1\over A^2 \Gamma_{\rm n} h} \unab^i \Psi \ . 
\ee 

By the means of Eq.~(\ref{e:kij}),
the momentum constraint equation yields
\be \label{e:shift}
\ulap N^i + {1 \over 3} \unab^i \left( \unab_j N^j \right) =
        - 16 \pi N A^2 (E + p) U^i + 2 N A^2 K^{ij} \unab_j (3 \beta - 4 \nu)
	\ . 
\ee

The equations to be solved to get the metric coefficients are the 
elliptic equations (\ref{e:beta}), (\ref{e:nu}) and
(\ref{e:shift}). Note that they 
represent only five of the ten Einstein equations. The remaining five Einstein
equations are not used in this procedure. Moreover, some of
these remaining equations
are certainly violated, reflecting the fact that the conformally flat 
3-metric (\ref{e:conf_plat}) is an approximation to the exact metric
generated by a binary system. 

At the Newtonian limit, Eqs.~(\ref{e:beta}) and (\ref{e:shift}) reduce 
to $0=0$. There remains only Eq.~(\ref{e:nu}), which gives the usual
Poisson equation for the gravitational potential $\nu$. 

\subsection{Equation for the matter}

Taking the logarithm of the first integral of motion (\ref{e:int_prem}) 
and using the metric (\ref{e:met}) yields
\be \label{e:int_prem_final}
  H + \nu + \ln\Gamma_{\rm n} + \ln\left( 1 + A^2 f_{ij} {B^i\over N} U^j
				 \right) = {\rm const} \ ,
\ee
where $H$ is the logarithm of the specific enthalpy:
\be
	H := \ln h \ . 
\ee
At the Newtonian limit, $H$ coincides with the specific non-relativistic
(i.e. which does not include the rest mass energy) enthalpy. 

Taking into 
account Eqs.~(\ref{e:gamma_n}) and (\ref{e:shift_non_rot}),
the Newtonian limit of the first integral (\ref{e:int_prem_final}) is
\be
	H + \nu + {1\over 2} (\nabla \Psi)^2
	- (\mib{\Omega} \times \mib{r})\cdot \nabla \Psi = {\rm const}.
\ee
(Recall that $\nu$ reduces to the Newtonian gravitational potential). 
We recognize the classical expression [compare e.g. with Eq.~(12) of 
Ref.~\citen{Teuko98} or Eq.~(11) of Ref.~\citen{UryuE98b}]. 

For a zero-temperature EOS, $H$ can be considered as a function of 
the baryon density $n$ solely, so that one can introduce the thermodynamical
coefficient
\be
	\zeta := {d\ln H \over d\ln n} \ .  
\ee
The gradient of $n$ which appears in Eq.~(\ref{e:psicov}) can be then
replaced by a gradient of $H$ so that, using the metric (\ref{e:conf_plat}),
one obtains
\be \label{e:psinum}
\zeta H \ulap \Psi + \unab^i H \unab_i \Psi =
  - A^2 h \Gamma_{\rm n} {B^i\over N} \unab_i H
 + \zeta H \left(
        \unab^i \Psi \unab_i(H-\beta) 
        - A^2 {B^i\over N} h \unab_i \Gamma_{\rm n}
                \right) \ .
\ee
Note that this is not a Poisson-type equation for $\Psi$, because the 
coefficient $H$ in front of the Laplacian operator vanishes at the
surface of the star. Numerically speaking, this means that this equation 
must be 
dealt by a specific technique and not by the direct use of 
some Poisson solver. 

At the Newtonian limit Eq.~(\ref{e:psinum}) reduces to 
\be
   \zeta H \Delta \Psi + \nabla H \cdot \nabla \Psi =
	(\mib{\Omega}\times\mib{r}) \cdot \nabla H \ .   
\ee
Here again, we recognize the classical expression [compare e.g. with 
Eq.~(13) of Ref.~\citen{Teuko98}]. 

\section{Analytical approach}
\label{s:analytic}

The above equations cannot be solved analytically, unless additional 
simplifying assumptions are introduced. Two such assumptions are that
\begin{enumerate}
\item the fluid is incompressible;
\item the 1-PN approximation to general relativity is used. 
\end{enumerate}
These assumptions have recently allowed Taniguchi\cite{Tanig99} 
to find analytical solutions
to the relativistic irrotational Darwin-Riemann problem. 
Note that at the 1-PN level, the figures are no longer ellipsoids, even 
for an incompressible fluid. This is explicitely taken into account in
Taniguchi's procedure \cite{Tanig99}, which improves over a
previous 1-PN study by Lombardi, Rasio \& Shapiro \cite{LombaRS97}, where
the figures where assumed to be ellipsoidal. 

The main findings of Taniguchi's study\cite{Tanig99} are that the orbital
separation at the innermost stable circular orbit (ISCO) 
\footnote{Taniguchi defines the ISCO as the location of the
energy minimum along a constant baryon number sequence of decreasing separation,
the true ISCO being certainly close to this point.}
is lower than in 
the Newtonian case, whereas the orbital angular at the ISCO is roughly the 
same than in the Newtonian case.  

\section{Numerical approaches} \label{s:num}

Recently three groups have obtained numerical solutions of the partial
differential equations presented in Sect.~\ref{s:conf_flat}, by the mean
of different numerical techniques. In chronological order, these groups
are
\begin{itemize}
\item our group \cite{BonazGM99a,BonazGM99c}, which has employed
a multi-domain spectral method \cite{BonazGM98a} with spherical coordinates;
\item Marronetti, Mathews \& Wilson \cite{MarroMW99a,MarroMW99b}, 
who have employed
single-domain finite differences with Cartesian coordinates;
\item Uryu \& Eriguchi \cite{UryuE99}, who have employed multi-domain finite
differences with spherical coordinates. 
\end{itemize}

In this Section, we discuss only the numerical technique developed in
our group, whereas in Sect.~\ref{s:num_res}, we will present the results
obtained by the three groups. 

\begin{figure}
\centering
\epsfig{figure=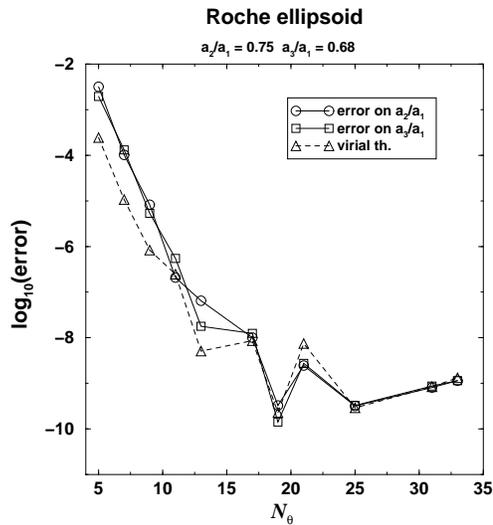,height=7cm}
\caption{Logarithm of the relative global error of the numerical solution
with respect to the number of degrees of freedom in
$\theta$ for a Roche ellipsoid for an equal mass binary system and
$\Omega^2/(\pi G\rho) = 0.1147$ (the numbers of degrees of freedom in the
other directions are 
$N_r = 2N_\theta-1$ and $N_\varphi = N_\theta -1$)\cite{BonazGM98a}. 
Also shown is the error in the
verification of the virial theorem.}
\label{f:test:roche}
\end{figure}

\subsection{Brief description of the multi-domain spectral method}

The numerical procedure to solve the PDE system presented in 
Sect.~\ref{s:conf_flat}
is based on the multi-domain spectral method developed in Ref.~\citen{BonazGM98a}.
We simply recall here some basic features of the method:
\begin{itemize}
\item Spherical-type coordinates $(\xi,\theta',\varphi')$ centered on each 
star are used: this ensures
a much better description of the stars  than by means of Cartesian coordinates. 
\item These spherical-type coordinates are surface-fitted coordinates: i.e.
the surface of each star lies at a constant value of the coordinate
$\xi$ thanks to a mapping $(\xi,\theta',\varphi')\mapsto (r,\theta,\varphi)$
[see Ref.~\citen{BonazGM98a} for details about this mapping]. This 
ensures that the spectral method applied in each domain is free from any
Gibbs phenomenon (spurious oscillations generated by discontinuities).  
\item The outermost domain extends up to spatial infinity, thanks to the
mapping $1/r = (1-\xi)/(2R_0)$. This enables us to put exact boundary conditions
on the elliptic equations (\ref{e:beta}), (\ref{e:nu}) and
(\ref{e:shift}) for the metric coefficients: spatial infinity
is the only location where the metric is known in advance (Minkowski metric). 
\item Thanks to the use of a spectral method \cite{BonazGM99b}
in each domain, the numerical
error is {\em evanescent}, i.e. it decreases exponentially with the number
of coefficients (or equivalently grid points) used in the spectral expansions,
as shown in Fig.~\ref{f:test:roche}.
\end{itemize}

\begin{figure}
\centering
\epsfig{figure=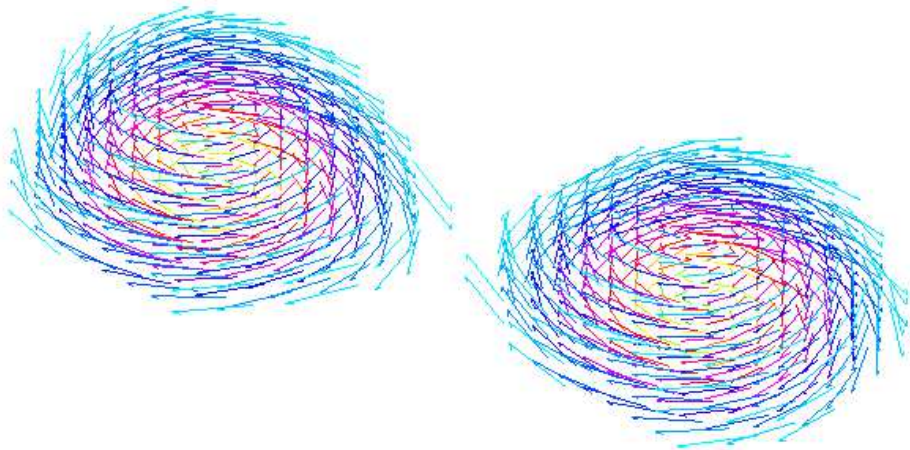,height=7cm}
\caption{Velocity field with respect to the co-orbiting frame in a
irrotational binary system (only the part of the stars above the orbital
plane is represented).}
\label{f:vitesse}
\end{figure}

The PDE system to be solved being non-linear, we use an iterative 
procedure. 
The iteration is stopped when the relative difference in the enthalpy field
between two successive steps goes below a certain threshold, typically
$10^{-7}$. An illustrative solution is shown 
in Fig.~\ref{f:vitesse}.

The numerical code is written in the {\sc LORENE} language 
\cite{MarckGG99}, which is a C++ based language for numerical relativity. 
A typical run make use of $N_r = 33$, $N_\theta = 21$, and
$N_\varphi=20$ coefficients (= number of
collocation points, which may be seen as number of grid points) in
each of the domains. 8 domains are
used : 3 for each star and 2 domains centered on the intersection between
the rotation axis and the orbital plane. The corresponding memory requirement
is 155 MB. A computation involves $\sim 250$ steps,
which takes 9~h~30~min on one CPU of a SGI Origin200 computer 
(MIPS~R10000 processor at 180 MHz). 
Note that due to the rather small memory requirement, 
runs can be performed in parallel on a multi-processor platform. 
This especially useful to compute sequences of configurations.  

\subsection{Tests passed by the code}

In the Newtonian and incompressible
limit, the analytical solution constituted by a Roche ellipsoid is
recovered with a relative accuracy of $\sim 10^{-9}$, as shown in  
Fig.~\ref{f:test:roche}. For compressible and irrotational Newtonian
binaries, no analytical solution is available, but the virial theorem
can be used to get an estimation of the numerical error: we found that the
virial theorem is satisfied with a relative accuracy of $10^{-7}$. 
Preliminary comparisons with the irrotational Newtonian configurations
recently computed by Uryu \& Eriguchi \cite{UryuE98a,UryuE98b} reveal
a good agreement.  Regarding the relativistic case, we
have checked our procedure of resolution of the gravitational field
equations by comparison with the results of Baumgarte et
al.~\cite{BaumgCSST98a} which deal with corotating binaries [our code
can compute corotating configurations by setting to zero the
velocity field of the fluid with respect to the co-orbiting observer]. 
We have performed the comparison with
the configuration $z_A=0.20$ in Table~V of Ref.~\citen{BaumgCSST98a}. We
used the same equation of state (EOS) (polytrope with $\gamma=2$), same
value of the separation $r_C$ and same value of the maximum density parameter
$q^{\rm max}$. We found a relative discrepancy of $1.1\%$ on $\Omega$,
$1.4\%$ on $M_0$, $1.1\%$ on $M$, $2.3\%$ on $J$, $0.8\%$ on $z_A$,
$0.4\%$ on $r_A$ and $0.07\%$ on $r_B$ (using the notations of
Ref.~\citen{BaumgCSST98a}).

\section{Numerical results} \label{s:num_res}

\subsection{Equation of state and compactification ratio}

As a simplified model for nuclear matter, 
let us consider a polytropic EOS with an adiabatic index
$\gamma=2$: 
\begin{equation} \label{e:eos}
p=\kappa (m_{\rm B} n)^\gamma \ , \qquad 
e=m_{\rm B} n + p/(\gamma-1) \ ,
\end{equation}
with $\kappa = 1.8 \times 10^{-2} {\ \rm J\, m}^3{\rm kg}^{-2}$. 
This EOS is the
same as that used by Mathews, Marronetti and Wilson
(Sect.~IV~A of Ref~\citen{MatheMW98}).

\begin{figure}
\centering
\epsfig{figure=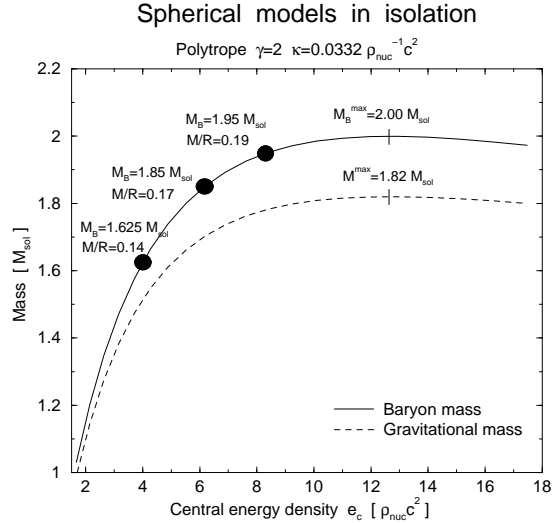,height=7cm}
\caption{Mass as a function of the central energy density for
static isolated neutron stars constructed with the EOS (\ref{e:eos}).
The heavy dots are configurations considered by our group 
\cite{BonazGM99a,BonazGM99c},
Marronetti et al. \cite{MatheMW98,MarroMW99a,MarroMW99b}, and
Uryu \& Eriguchi \cite{UryuE99}
(see text)
($\rho_{\rm nuc} := 1.66\times 10^{17} {\rm \ kg\, m}^{-3}$).}
\label{f:m_sher}
\end{figure}

The mass -- central density curve of static
configurations in isolation constructed upon this EOS is represented
in Fig.~\ref{f:m_sher}. The three points on this curve corresponds to 
three configurations studied by the various groups mentioned in the 
beginning of Sect.~\ref{s:num}:
\begin{itemize}
\item The configuration of baryon mass $M_{\rm B} = 1.625 \, M_\odot$ 
and compactification ratio $M/R = 0.14$ is
that considered in the dynamical study of Mathews, Marronetti and Wilson
\cite{MatheMW98} and in the quasiequilibrium studies of our group 
\cite{BonazGM99a,BonazGM99c}, of Marronetti et al. \cite{MarroMW99b}, and
of Uryu \& Eriguchi \cite{UryuE99}. 
\item The configuration of baryon mass $M_{\rm B} = 1.85 \, M_\odot$ 
and compactification ratio $M/R = 0.17$ has been studied 
by our group \cite{BonazGM99c} and Uryu \& Eriguchi \cite{UryuE99}.
\item The configuration of baryon mass $M_{\rm B} = 1.95 \, M_\odot$ 
and compactification ratio $M/R = 0.19$ has been studied  by
Marronetti et al.\cite{MarroMW99a,MarroMW99b}
\footnote{Marronetti et al.
\cite{MarroMW99a,MarroMW99b} use a different value for the EOS constant $\kappa$:
their baryon mass $M_{\rm B} = 1.55 \, M_\odot$ must be rescaled to our
value of $\kappa$ in order to get $M_{\rm B} = 1.95 \, M_\odot$. Apart from
this scaling, this is the same configuration, i.e. it has the same 
compactification ratio $M/R = 0.19$ and its relative distance with respect to 
the maximum mass configuration, as shown in Fig.~\ref{f:m_sher}, is the same.}.
\end{itemize}

\subsection{Irrotational sequence with $M/R=0.14$}

\begin{figure}
\centering
\epsfig{figure=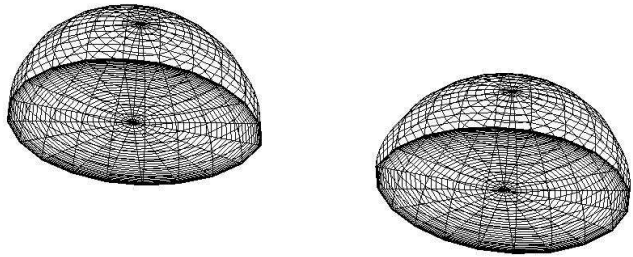,height=7cm}
\caption{Shape of irrotational binary neutron stars of baryon mass 
$M_{\rm B} = 1.625 \, M_\odot$, when the coordinate separation between
their centers (density maxima) is $41{\rm\ km}$. Only one half of each star
is represented (the part which is above the orbital plane).
The drawing is that of the numerical grid, which coincides with the surface
of the star, thanks to the use of surface-fitted spherical coordinates.}
\label{f:surf-3D}
\end{figure}

In this section, we give some details about the irrotational sequence
$M_{\rm B} = 1.625 \, M_\odot$ presented in Ref.~\citen{BonazGM99a}.
This sequence starts at the coordinate separation $d=110 {\rm\ km}$ (orbital
frequency $f=82{\rm\ Hz}$), where the two stars are almost spherical, and
ends at $d=41 {\rm\ km}$ ($f=332{\rm\ Hz}$), where a cusp appears on the
surface of the stars, which means that the stars start to break. 
The shape of the surface at this last point is shown in Fig.~\ref{f:surf-3D}.

\begin{figure}
\centering
\epsfig{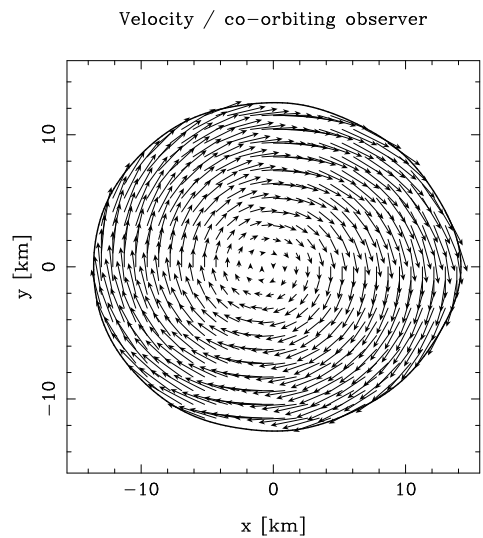}
\caption{Velocity field with respect to the co-orbiting observer, for the
configuration shown in Fig.~\ref{f:surf-3D}. The plane of the figure is the
orbital plane. The heavy line denotes the surface of the star.
The companion is located at $x=+41{\rm\ km}$.}
\label{f:vit}
\end{figure}

The velocity field with respect to the co-orbiting observer, as defined 
by Eq.~(52) of Ref.~\citen{BonazGM97b}, is shown in Fig.~\ref{f:vit}. 
Note that this field is tangent to the surface of the star, as it must be.

\begin{figure}
\centering
\epsfig{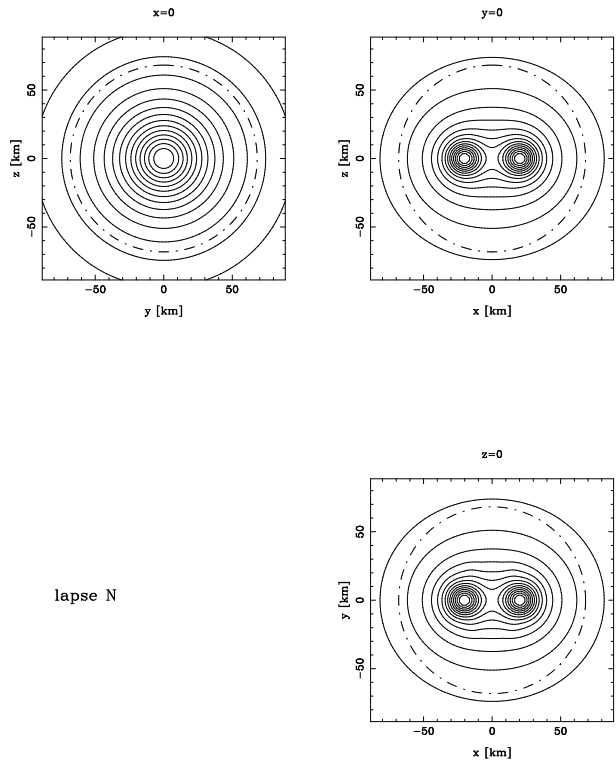}
\caption{Isocontour of the lapse function $N$ for the
configuration shown in Fig.~\ref{f:surf-3D}. The plots are cross section
in the $x=0$, $y=0$ and $z=0$ planes (note that the $x$ coordinate is
shifted by $20.5{\rm\ km}$ with respect to that of Fig.~\ref{f:vit}).
The dot-dashed line denotes the
boundary between the inner numerical grid and the outer compactified one 
(which extends to spatial infinity), for the grid system centered on
the intersection between the rotation axis and the orbital plane.}
\label{f:lapse}
\end{figure}

\begin{figure}
\centering
\epsfig{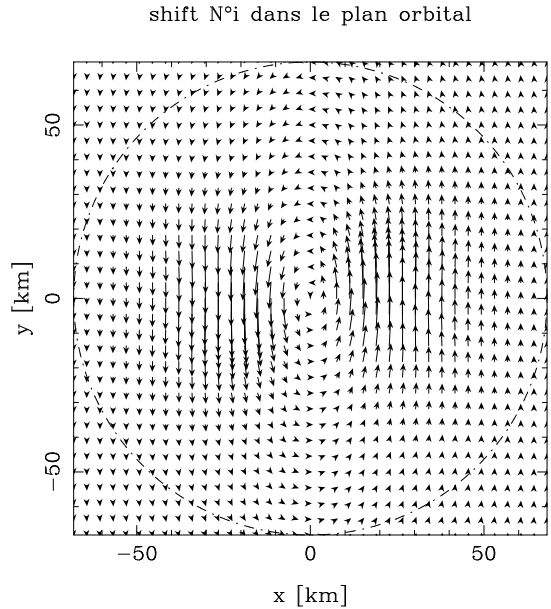}
\caption{Shift vector of non-rotating coordinates in the orbital plane, 
for the configuration shown in Figs.~\ref{f:surf-3D}-\ref{f:lapse}.}
\label{f:shift}
\end{figure}

The lapse function $N$ is represented in 
Fig.~\ref{f:lapse}. The coordinate system $(x,y,z)$ is centered on
the intersection between the rotation axis and the orbital plane.
The $x$ axis joins the two stellar centers, and $z=0$ is the orbital
plane. The value of $N$ at the center of each star is
$N_{\rm c}=0.64$, whereas the central value of the conformal factor $A^2$ is
$A^2_{\rm c}=2.20$. 
The shift vector of non-rotating coordinates, $\bf N$, 
is shown in Fig.~\ref{f:shift}.
Its maximum value is $0.10\, c$. 

\begin{figure}
\centering
\epsfig{figure=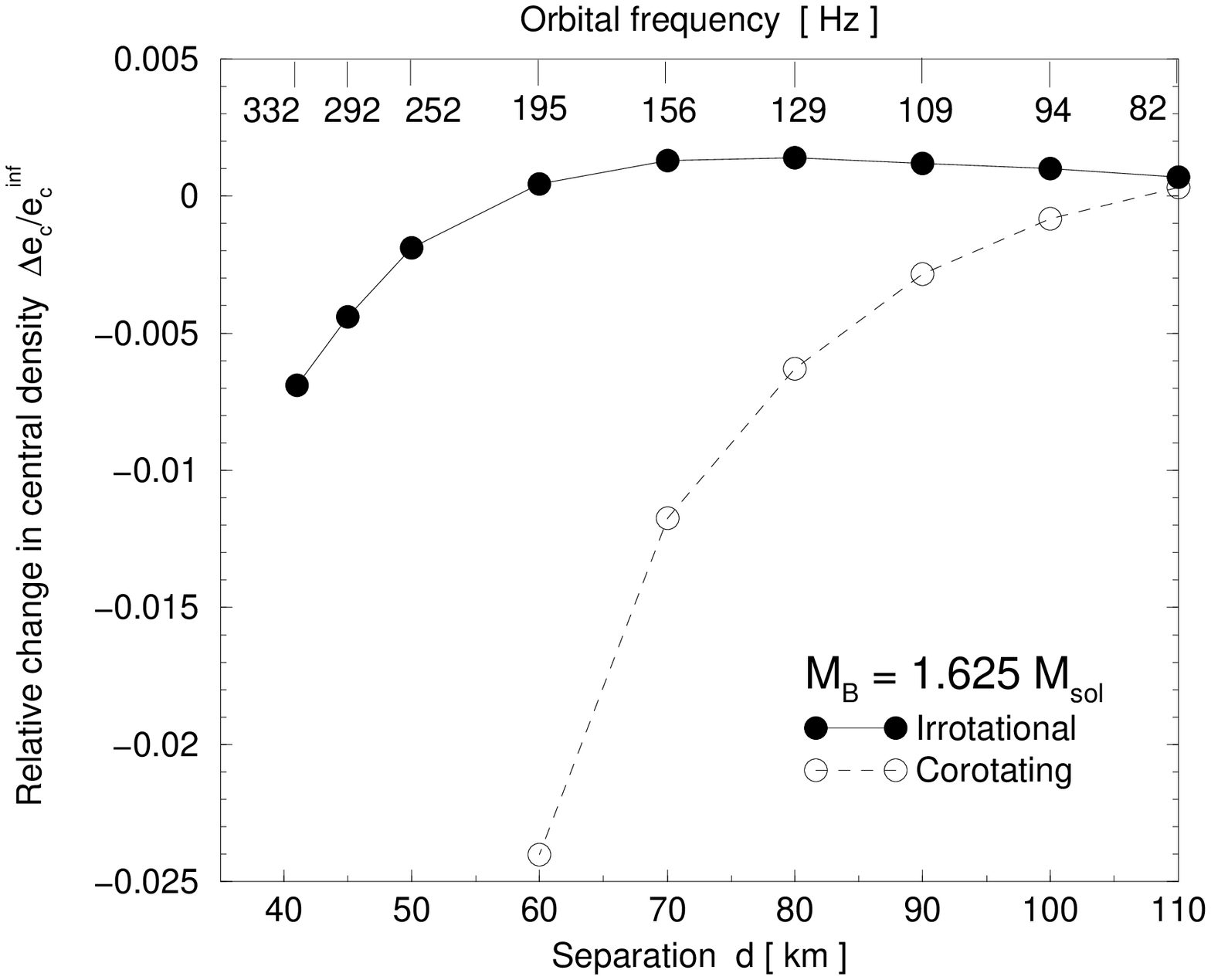,height=7cm}
\caption{Relative variation of the central energy density $e_{\rm c}$ with
respect to its value at infinite separation $e_{\rm c}^{\rm inf}$ as a
function of the coordinate separation $d$ (or of the orbital frequency
$\Omega/(2\pi)$) for constant baryon mass $M_{\rm B} = 1.625 \,
M_\odot$ sequences\cite{BonazGM99a}.  
The solid (resp. dashed) line corresponds to a
irrotational (resp. corotating) sequence of coalescing neutron star
binaries.}
\label{f:cent_dens_1.625}
\end{figure}

The variation of the central density along the $M_{\rm B}=1.625\, M_\odot$
sequence is shown in Fig.~\ref{f:cent_dens_1.625}. We have also
computed a corotating (i.e. synchronized) 
sequence for comparison (dashed line in 
Fig.~\ref{f:cent_dens_1.625}). In the corotating case, the central density
decreases quite substantially as the two stars approach each other. 
This is in agreement with the results of Baumgarte et 
al.~\cite{BaumgCSST97,BaumgCSST98a}. In the irrotational case (solid
line in Fig.~\ref{f:cent_dens_1.625}), the central density remains rather
constant (with a slight increase, below $0.1\%$)
before decreasing. 
This contrasts with the result of the dynamical calculations by 
Mathews et al.~\cite{MatheMW98} which showed a central density increase
of $14\%$ for the same compactification ratio $M/R=0.14$. But, as stated
in Sect.~\ref{s:intro}, this last result has revealed to be spurious
due to some error in a formula employed in the code \cite{Flana99}.

It is worth to note that, for the same compactification ratio $M/R=0.14$, 
the computations by the other groups
show a central density variation in accordance with the one presented above:
it is at most 1\% both in the results of Marronetti et al. \cite{MarroMW99b}
and Uryu \& Eriguchi \cite{UryuE99}.

We can thus conclude that no tendency to individual
gravitational collapse is found when the orbit shrinks.  

\subsection{Irrotational sequence with $M/R=0.17$}

\begin{figure}
\centering
\epsfig{figure=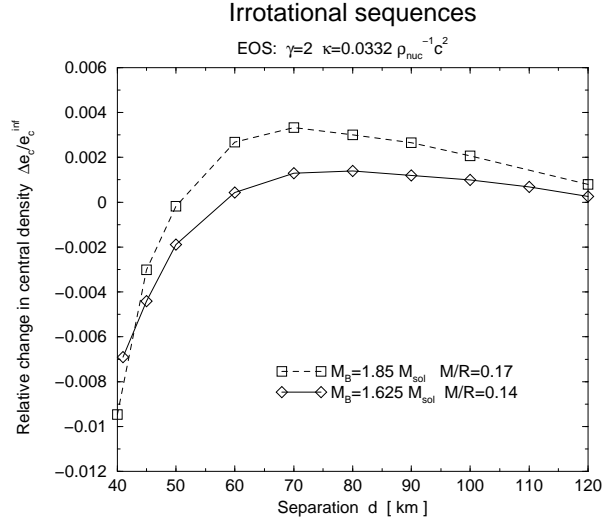,height=7cm}
\caption{Relative variation of the central energy density $e_{\rm c}$ with
respect to its value at infinite separation $e_{\rm c}^{\rm inf}$ as a
function of the coordinate separation $d$ for constant baryon mass 
sequences with 
$M_{\rm B} = 1.625 \, M_\odot$ (solid line, same as in 
Fig.~\ref{f:cent_dens_1.625}) and
$M_{\rm B} = 1.85\, M_\odot$ (dashed line).}
\label{f:cent_dens_1.625_1.85}
\end{figure}

In order to investigate the dependency of the above result 
on the compactness
of the stars, we have computed an irrotational sequence with a baryon
mass $M_{\rm B} = 1.85\, M_\odot$, which corresponds to a compactification
ratio $M/R=0.17$ for stars at infinite separation
(second heavy dot in Fig.~\ref{f:m_sher}) \cite{BonazGM99c}. 
The result is compared
with that of $M/R=0.14$ in Fig.~\ref{f:cent_dens_1.625_1.85}. 
A very small density increase (at most $0.3\%$) 
is observed before the decrease. 
For the same compactification ratio $M/R=0.17$, Uryu \& Eriguchi \cite{UryuE99}
report a slightly higher increase (1.5 \%). 

This density increase remains
within the expected error ($\sim 2\%$, cf. Sect.~\ref{s:justif}) induced
by the conformally flat approximation for the 3-metric, so that it cannot
be asserted that this effect would remain in a complete calculation.

\subsection{Irrotational sequence with $M/R=0.19$}

Marronetti, Mathews and Wilson~\cite{MarroMW99a,MarroMW99b} 
have computed
quasiequilibrium irrotational configurations with a
compactification ratio $M/R=0.19$ (third heavy dot in Fig.~\ref{f:m_sher}).
They found a central density increase, as the orbit shrinks, of the order
of $1.5\%$. This still remains within the error introduced by the
conformally flat approximation.  
 
\section{Innermost stable circular orbit} \label{s:ISCO}

An important parameter for the detection of a gravitational wave signal from
coalescing binaries is the location of the innermost stable circular 
orbit (ISCO), if any. In Table~\ref{t:isco}, we recall what is known about
the existence of an ISCO for extended fluid bodies. The case of two point
masses is discussed in details in Ref.~\cite{BuonaD99}.

\begin{table}[t]
\begin{center}
\begin{tabular}{*{3}{c}}
\hline
\\[0.5ex]
Model & Existence of an ISCO & References \\[0.5ex]
\hline
\\[0.5ex]
Newtonian corotating & ISCO $\Leftrightarrow \, \gamma > 2$ & \citen{ShibaTN97}
\\[0.5ex]
Newtonian irrotational & ISCO $\Leftrightarrow \, \gamma > 2.4$ &
\citen{UryuE98a} 
\\[0.5ex]
GR corotating & ISCO $\Leftrightarrow \, \gamma > 5/3$ &
\citen{BaumgCSST97}, \citen{ShibaTN97} 
\\[0.5ex]
GR irrotational & ISCO exists for $\gamma = \infty$ &
\citen{Tanig99} 
\\[0.5ex]
\hline
\end{tabular}
\vspace{3mm}
\caption{Known results about the existence of an ISCO for extended fluid
bodies, in terms of the adiabatic index $\gamma$.}
\label{t:isco}
\end{center}
\end{table}

\begin{figure}
	\parbox{\halftext}{
		\epsfig{figure=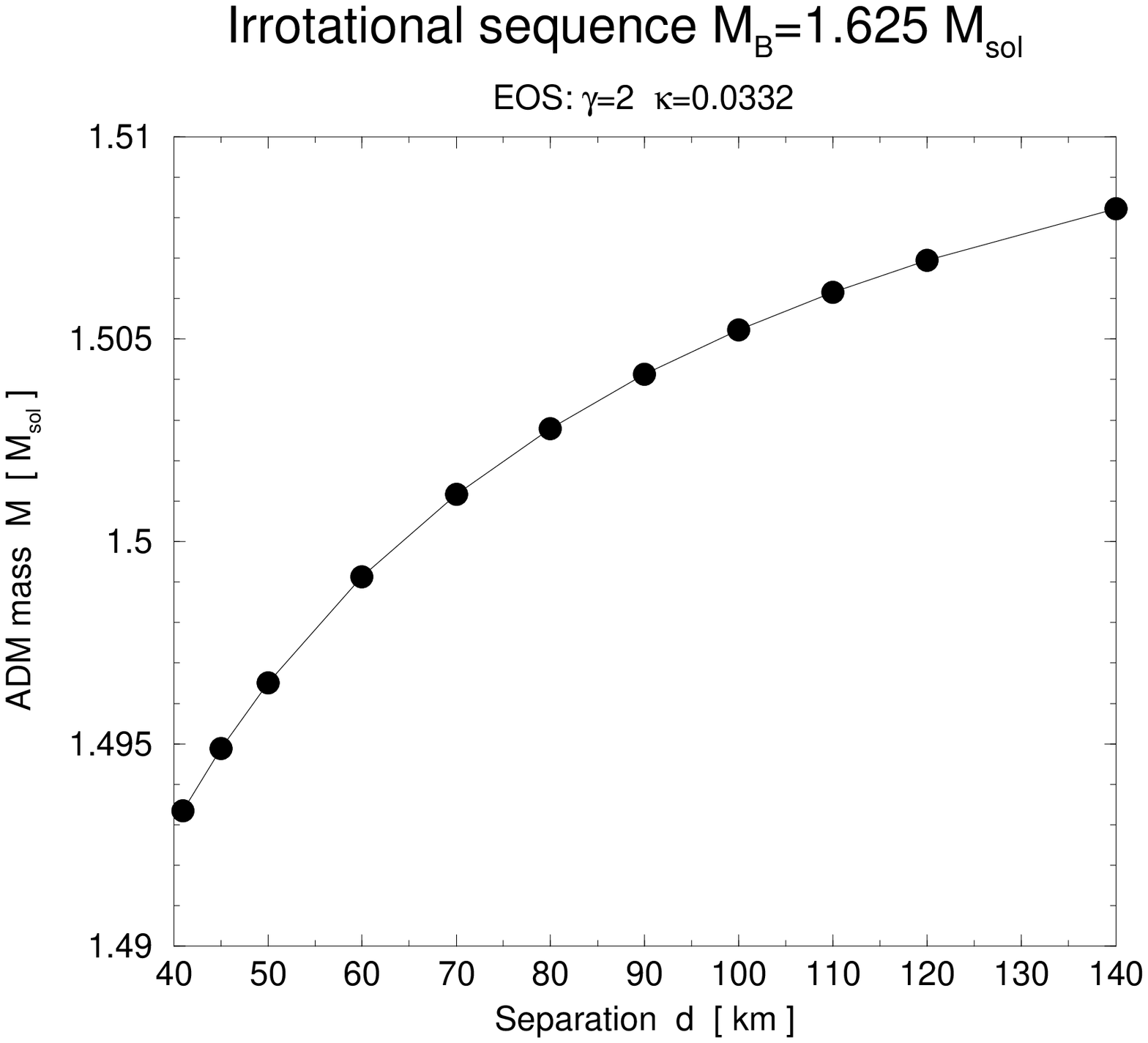,height=5cm}}
	\hspace{8mm}
	\parbox{\halftext}{
		\epsfig{figure=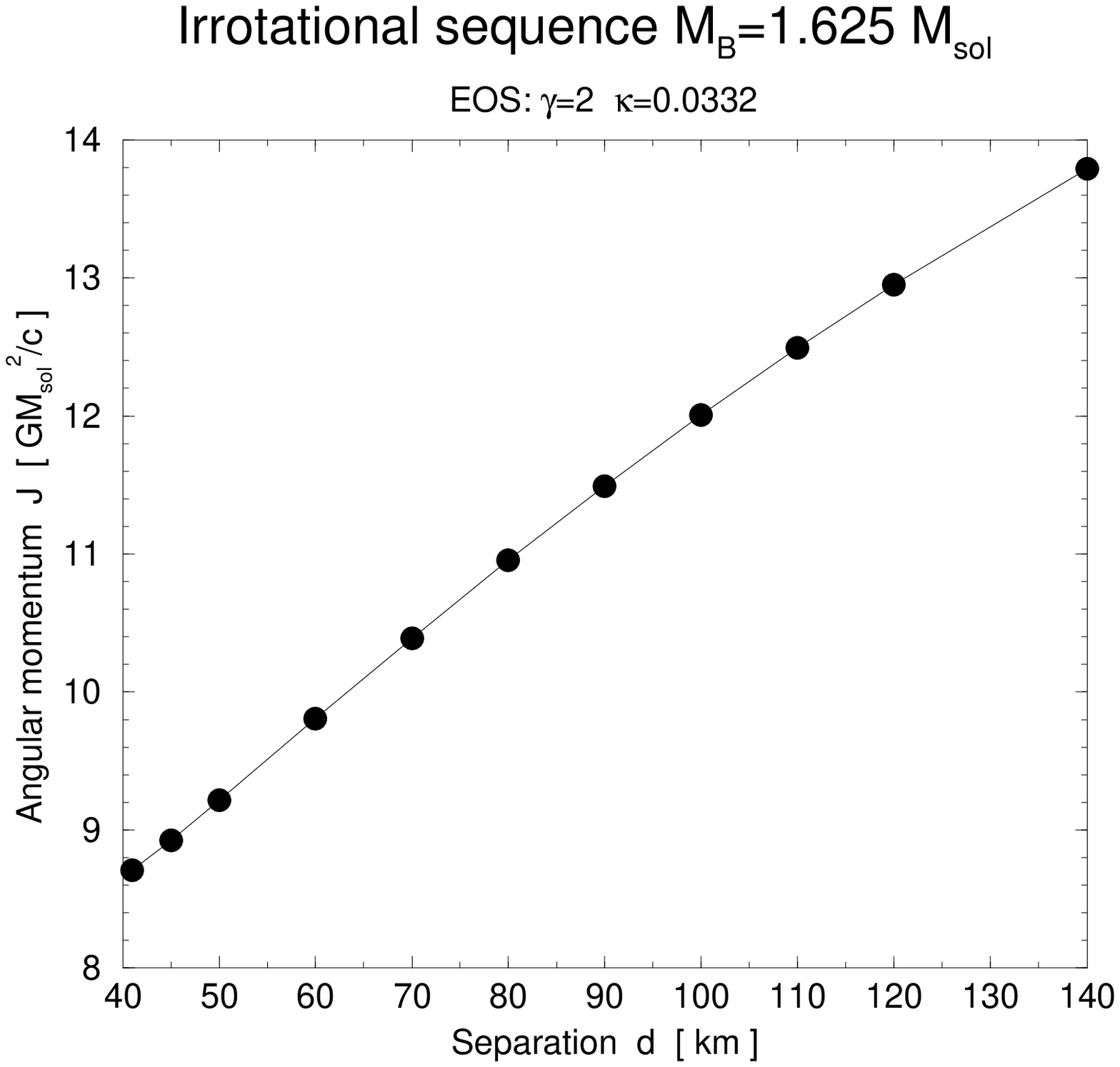,height=5cm}}
	\parbox{\halftext}{
		\epsfig{figure=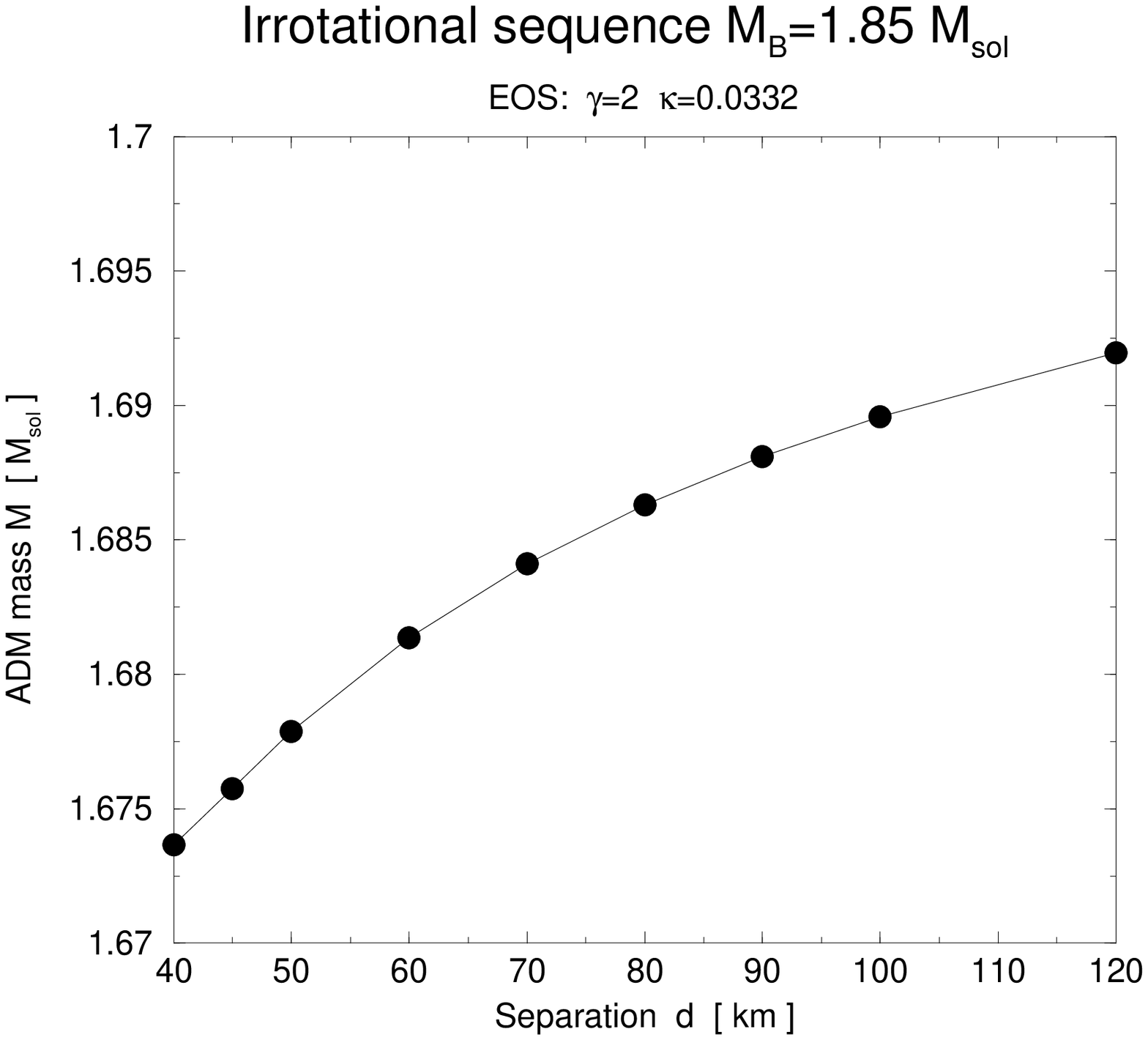,height=5cm}}
	\hspace{8mm}
	\parbox{\halftext}{
		\epsfig{figure=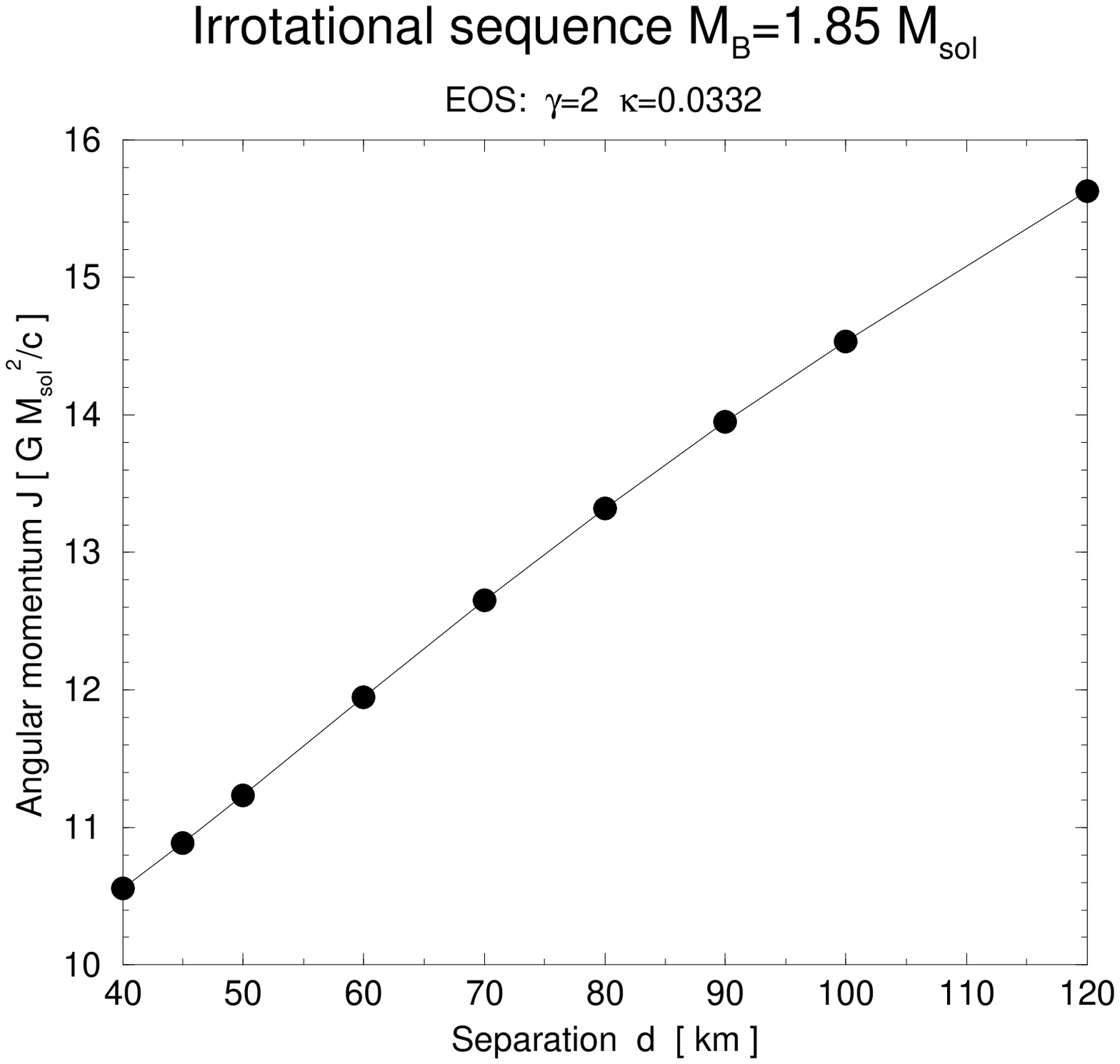,height=5cm}} 
\caption{Variation of (half of) the ADM mass (left) and of the total
angular momentum (right) of the binary system with respect to the
coordinate distance $d$, along the evolutionary sequences 
$M_{\rm B} =  1.625\, M_\odot$, $M/R=0.14$ (top) and 
$M_{\rm B} =  1.85\, M_\odot$,  $M/R=0.17$ (bottom) 
[from Ref.~\cite{BonazGM99c}]}.
\label{f:mj_d} 
\end{figure}

For Newtonian binaries, it has been shown \cite{LaiRS93} that the ISCO
is located at a minimum of the total energy, as well as total angular momentum,
along a sequence at constant baryon number and constant circulation 
(irrotational sequences are such sequences). The instability found in this
way is dynamical. For corotating sequences, it is secular instead 
\cite{LaiRS93,LaiRS94}.
This turning point method also holds for locating ISCO in 
relativistic corotating binaries \cite{BaumgCSST98b}. 
For relativistic irrotational configurations, no rigorous theorem has been
proven yet about the localization of the ISCO by a turning point method. 
All what can be said is that no turning point is present in the 
irrotational sequences considered in Sect.~\ref{s:num_res}:
Fig.~\ref{f:mj_d} shows the variation as the orbit shrinks 
of the ADM mass of the spatial hypersurface $t={\rm const}$ (which is a 
measure of the total energy, or equivalently of the
binding energy, of the system), as well as of the total angular momentum,
for the evolutionary sequences with $M/R=0.14$ and $M/R=0.17$. 
Clearly, both the ADM mass and the angular momentum decreases
monoticaly, without showing any turning point. 
The same result has been found by Marronetti et al. [Fig.~2 of
Ref.~\citen{MarroMW99b}, which also shows the good agreement between
Marronetti et al.\cite{MarroMW99b} results and ours\cite{BonazGM99c}]
and Uryu \& Eriguchi [Fig.~5 of Ref.~\citen{UryuE99}]. 

\section*{Acknowledgements}
We warmly thank Jean-Pierre Lasota for his constant support and
Brandon Carter for illuminating discussions.

\end{document}